\documentclass[letter]{aa}
\usepackage[dvips]{graphicx}
\usepackage{txfonts}
\usepackage{natbib}
\bibpunct{(}{)}{;}{a}{}{,}

\begin{document}

\title{RXTE confirmation of the intermediate polar status of IGR~J15094--6649}

\author{O.W. Butters\inst{1,2}
\and
  A.J. Norton\inst{2}
\and
  K. Mukai\inst{3,4}
\and
  E.J. Barlow\inst{2}
}

\institute{
  Department of Physics and Astronomy, University of Leicester,
  Leicester, LE1 7RH, UK\\
  \email{oliver.butters@star.le.ac.uk}
\and
  Department of Physics and Astronomy, The Open University, Walton
  Hall, Milton Keynes MK7 6AA, UK
\and
 CRESST and X-ray Astrophysics Laboratory NASA/GSFC, Greenbelt,
        MD 20771, USA
\and
 Department of Physics, University of Maryland, Baltimore County, 1000
  Hilltop Circle, Baltimore, MD 21250, USA.
}

\authorrunning{Butters et al.}

\date{Accepted 2009 ???;
      Received  2009 ???;
      in original form 2009 ???}

\abstract
{}
{To establish the X-ray properties of the intermediate polar candidate
 IGR~J15094--6649 and therefore confirm its inclusion into the class.}
{42\,856~s of X-ray data from \textit{RXTE} was analysed.
Frequency analysis was used to constrain temporal variations and
spectral analysis used to characterise the emission and absorption properties.}
{A spin period of 809.7$\pm$0.6~s is present, revealed as a complex pulse profile whose modulation depth decreases with increasing X-ray energy. The spectrum is well fitted
by either a 19$\pm$4~keV Bremsstrahlung or $\Gamma$=1.8$\pm$0.1 power
law, with an iron emission line feature and significant absorption in each case.}
{IGR~J15094--6649 is confirmed to be an intermediate polar.}


\maketitle

\section{Introduction}

Intermediate polars (IPs) are members of the cataclysmic variables
class. They are thought to harbour a magnetic field strong
enough to greatly influence the accretion from the main sequence star
to the white dwarf, but not strong enough to synchronise the
system. For this situation to occur the magnetic field is believed to
to be in the range of a few MG to tens of MG at the white
dwarf surface. This has the effect of channelling the accreting
material onto the magnetic poles of the white dwarf and causes a hot
dense accretion column to form, which emits high energy X-rays. As the
white dwarf spins, the absorbing column density in the line of sight varies;
this gives rise to one of the defining characteristics of IPs -- X-ray modulation
at the spin period with a modulation depth which decreases as the X-ray energy
increases. For an exhaustive review of CVs see e.g. \cite{warner95}.

The exact number of IPs is heavily dependent on the selection criteria
used, but the lower end estimate is currently taken to be thirty
three\footnote{http://asd.gsfc.nasa.gov/Koji.Mukai/iphome/iphome.html
  (IP catalogue version 2008b)}.

In recent years hard X-ray telescopes ({\sl INTEGRAL} and Swift) have
unexpectedly found many known IPs and discovered several candidate systems.
This has raised the question of whether a sample of hard X-ray selected candidate
IPs would be different from the current soft X-ray selected population. With this in
mind the candidate \object{IGR~J15094--6649} (hereafter J1509) is studied. This
forms part of an ongoing survey to classify hard X-ray selected IPs
\citep{butters07,butters08}.

\section{Previous observations of J1509}

In the {\sl INTEGRAL}/IBIS survey J1509 was detected as an unclassified object
in the 20-100~keV energy range \citep{barlow06}. Both a Bremsstrahlung and a power law
were fitted to the data in order to determine an identification. In both cases a
good fit was found; Bremsstrahlung with kT=13.8$\pm$5.1~keV and power law
with $\Gamma$=3.6$\pm$0.8. The flux was given as
1.38$\times$10$^{-11}$~ergs~s$^{-1}$~cm$^{-2}$ in the 20--100~keV band.

\citet{masetti06} classified J1509 as an IP based upon optical spectra taken
at the 1.5~m Cerro Tololo Interamerican Observatory (CTIO) in Chile.

Very recently \citet{pretorius09} published optical photometry and spectroscopy
of J1509, along with four other candidate IPs from the {\em INTEGRAL} sample. She
detected a clear radial velocity signal at a period of $5.89 \pm 0.01$~h which was
identified as the orbital period of the system, as well as a photometric modulation
at $809.42 \pm 0.02$~s which was taken to be the spin period of the magnetic white
dwarf. These results provided very strong indications that J1509 is an IP, and
detection of a commensurate pulse period in X-ray data would absolutely confirm
its classification.

\section{Observations and data reduction}

Data were obtained from the {\em RXTE} satellite \citep{bradt93} with the PCA
instrument over two consecutive days, from 30th -- 31st December 2008. The total time
on target was 42\,856~s. Initial data reduction was
done with the standard {\sc ftools}. Only the top layer of PCU2 was included in
the measurements and we used the standard 2 mode data with a time
resolution of 16~s. Background
subtracted light curves were constructed in four
energy bands: 2~--~4~keV, 4~--~6~keV, 6~--~10~keV and 10~--~20~keV, as well as
a combined 2~--~10~keV band for maximum signal-to-noise. A mean X-ray
spectrum was also extracted.

\subsection{Light curve analysis}

\begin{figure}
  \resizebox{\hsize}{!}{\includegraphics{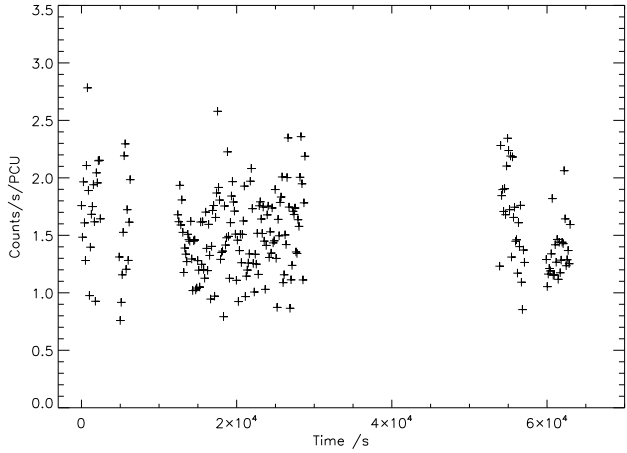}}
  \caption{2~--~10~keV background subtracted light curve of J1509. The zero time
  corresponds to the start of the observations at JD2454830.77312937984. The
  data is binned into bins of 128~s width. The typical error on each
  point is 0.26 counts~s$^{-1}$~PCU$^{-1}$.}
  \label{unfolded}
\end{figure}

\begin{figure}
  \resizebox{\hsize}{!}{\includegraphics{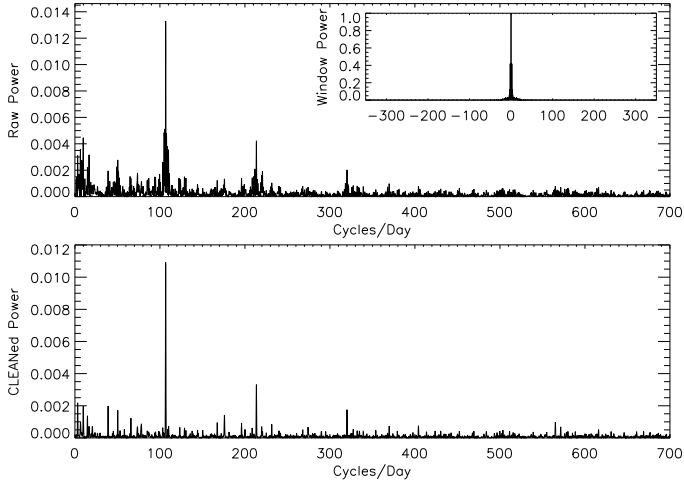}}
  \caption{2-10~keV {\sc clean}ed periodogram. The upper plot shows the raw periodogram,
  with the window function inset; the lower plot shows the deconvolved
  ({\sc clean}ed) periodogram.}
  \label{cleaned}
\end{figure}

In the 2~--~10~keV energy band the raw count rate varied between 2.5
  and 6.9~count~s$^{-1}$~PCU$^{-1}$. The background count rate,
  generated using the faint source background model, varied between 2.9 and
  4.1~count~s$^{-1}$~PCU$^{-1}$. The background subtracted 2~--~10~keV light curve is shown in Fig.~\ref{unfolded}. The data were subsequently analysed with a variable gain implementation of the {\sc clean} algorithm \citep{lehto97}
to discover any periodicities and discount any aliasing effects. The results of
this are shown in Fig.~\ref{cleaned}.

\begin{figure}
  \resizebox{\hsize}{!}{\includegraphics{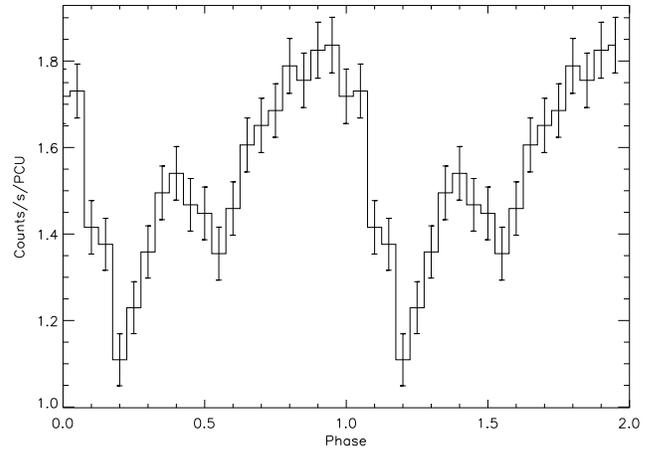}}
  \caption{2~--~10~keV light curve folded at the 809.7~s period
  with an arbitrary zero point. Two cycles are shown for clarity.}
  \label{spin_folded}
\end{figure}

A strong peak is evident in the {\sc clean}ed periodogram at
approximately 107~cycles~day$^{-1}$, in the 2~--~10~keV energy
band. Also present are its first and second harmonics at $\sim~214$
and $\sim~321$ cycles~day$^{-1}$ respectively. Analysis of the second harmonic
peak yields a fundamental pulsation period of $809.7\pm0.6$~s (based on a
Gaussian fit to the periodogram). This is in excellent agreement with the optical
photometric period detected by \citet{pretorius09}. Each of the energy resolved light
curves were folded at the 809.7~s period, and Fig.~\ref{spin_folded} shows the result in
the 2~--~10~keV energy band. The modulation depths of the pulse profile
were then estimated by fitting a sinusoid to the folded data in each
energy band and dividing the semi-amplitude by the fitted mean. The
results of this are shown in Table~\ref{modulation_depths}.

\begin{table}
  \caption{Modulation depths of the pulse profile in different  energy bands.
  Modulation depth is defined here as the semi-amplitude of a fitted sinusoid
  divided by the fitted mean.}
  \label{modulation_depths}
  \centering
  \begin{tabular}{ccc}
    \hline\hline
    Energy band & Modulation depth & Fitted mean\\
    (keV)       & (\%)             & (ct~s$^{-1}$~PCU$^{-1}$)\\
    \hline
    2--10       & 15$\pm$1 & 1.54\\
    2--4        & 27$\pm$3 & 0.33\\
    4--6        & 16$\pm$2 & 0.54\\
    6--10       & 9$\pm$2  & 0.67\\
    10--20      & 7$\pm$4  & 0.34\\
    \hline
  \end{tabular}
\end{table}

\begin{figure}
  \resizebox{\hsize}{!}{\includegraphics{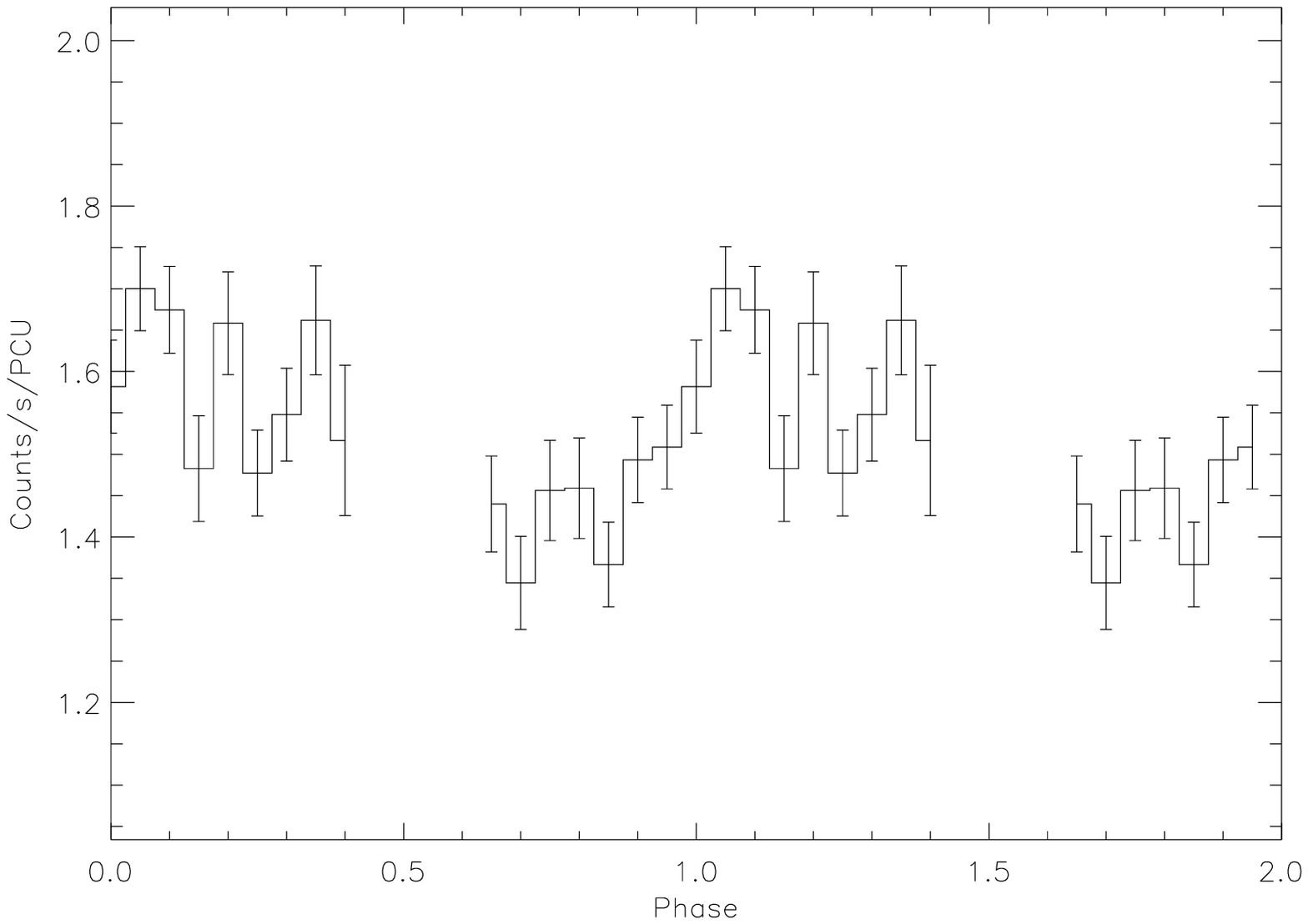}}
  \caption{2~--~10~keV light curve folded at the 21\,204~s period of
  \citet{pretorius09} with an arbitrary zero point. Two cycles are shown
  for clarity.}
  \label{orb_folded}
\end{figure}

\begin{table*}
  \centering
  \caption{Bremsstrahlung (top) and power law(bottom) spectral fitting
  parameters of J1509. $n_{\rm H}$(Galactic)=$0.2~\times
  10^{22}$~cm$^{-2}$.}
  \label{spectral_fits}
  \centering
  \begin{tabular}{cccccccc}
    \hline\hline
    $n_{\rm H}$         & $kT$     & $\Gamma$    & Fe           & $\sigma_{\rm Fe}$ & EW   & $\chi^2_{\rm reduced}$ & Flux (2--10keV)\\
    10$^{22}$~cm$^{-2}$ & keV      &             & keV          & keV               & keV  &                        & 10$^{-11}$~ergs~cm$^{-2}$~s$^{-1}$\\
    \hline
    1.2$\pm$1.0         & 19$\pm$4 & --          & 6.3$\pm$0.1  & 0.4$\pm$0.2       & 0.9  & 0.6                    & 1.6 \\
    3.0$\pm$1.3         & --       & 1.8$\pm$0.1 & 6.4$\pm$0.1  & 0.4$\pm$0.2       & 0.9  & 0.6                    & 1.5 \\
    \hline
  \end{tabular}
\end{table*}

There is no indication in the power spectrum of the spectroscopic
orbital period previously reported by \citet{pretorius09}. Folding the
X-ray light curve at the proposed orbital period yields a profile with
no significant coherent modulation (see Fig.~\ref{orb_folded}). 

\subsection{Spectral Analysis}

\begin{figure}
  \resizebox{\hsize}{!}{\rotatebox{-90}{\includegraphics{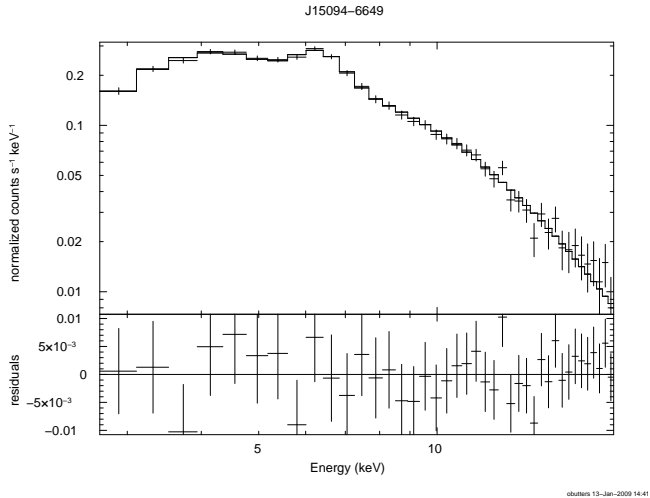}}}
  \caption{2.5~--~20 keV mean spectrum fitted with a photoelectrically
  absorbed Bremsstrahlung plus iron line profile.}
  \label{brems_spectrum}
\end{figure}

Analysis of the X-ray spectrum was carried out with the {\sc xspec}
package. Two models were used for fitting; a photoelectrically
absorbed Bremsstrahlung, and a photoelectrically absorbed power
law. Both models had an excess at approximately 6.4~keV, so a Gaussian
was added to account for the iron line emission. Both models gave a good
fit to the data (see Fig.~\ref{brems_spectrum} and
Table~\ref{spectral_fits}).

\section{Discussion}

The X-ray period found here (809.7~s) is in agreement with the photometric
pulsation period found by \citet{pretorius09} and is typical of a white dwarf
spin period in an IP. Furthermore,  the
increasing modulation depth with decreasing energy in the folded pulse
profiles is an indication that an accretion column absorbing structure is present
\citep{norton89}. The complex pulse profile and presence of strong
harmonics in the power spectrum are reminiscent of the canonical IP
\object{FO~Aqr} \citep{beardmore98}, although unlike that system, there is here
no evidence for an additional X-ray modulation at the beat period
(841.5~s), which would be indicative of a stream-fed component to the
accretion. We note that whilst some IPs exhibit the white dwarf spin
period in their X-ray flux, they may show the beat period in optical
photometry (for example, \object{AO~Psc}). This arises
due to reprocessing of the X-ray signal, probably from the face of the
donor star. In the case of J1509 we can be confident that we are
seeing the spin period of the white dwarf in both the optical and
X-ray light curves. The length of the X-ray data set probably precludes the detection of the orbital period, or may
indicate that the system is seen at relatively low inclination angle
\citep{parker05}, so no such modulation is present.

Both spectral fits are good and the Bremsstrahlung model in particular is in agreement
with that seen in the {\em INTEGRAL} data at higher energies \citep{barlow06}. The fit
parameters are typical of those seen in other IPs. The column density
is greater than the Galactic column density (as given by the HEASARC
$n_{\rm H}$ estimator\footnote[1]{http://heasarc.nasa.gov/cgi-bin/Tools/w3nh/w3nh.pl}).
This too is typical of IPs and is likely due to absorption by material
within the accretion flow.
The {\em ROSAT} Bright Source Catalogue has one other source in the
{\em RXTE} field of view. A count rate for it was estimated using the
webPIMMS\footnote[2]{http://www.ledas.ac.uk/pimms/w3p/w3pimms.html} tool
and scaled according to the response of the detector. The count rate of this
additional source was small ($\sim~0.04$~counts~s$^{-1}$~PCU$^{-1}$) and
therefore does not affect our result.

\section{Conclusion}
\object{IGR~J15094--6649} is confirmed as an IP and adds to the growing list of
hard X-ray selected magnetic CVs discovered by {\em INTEGRAL}.

\bibliographystyle{aa}
\bibliography{ref}

\begin{thebibliography}{11}
\expandafter\ifx\csname natexlab\endcsname\relax\def\natexlab#1{#1}\fi

\bibitem[{{Barlow} {et~al.}(2006){Barlow}, {Knigge}, {Bird}, {J Dean}, {Clark},
  {Hill}, {Molina}, \& {Sguera}}]{barlow06}
{Barlow}, E.~J., {Knigge}, C., {Bird}, A.~J., {et~al.} 2006, MNRAS, 372, 224

\bibitem[{{Beardmore} {et~al.}(1998){Beardmore}, {Mukai}, {Norton}, {Osborne},
  \& {Hellier}}]{beardmore98}
{Beardmore}, A.~P., {Mukai}, K., {Norton}, A.~J., {Osborne}, J.~P., \&
  {Hellier}, C. 1998, MNRAS, 297, 337

\bibitem[{{Bradt} {et~al.}(1993){Bradt}, {Rothschild}, \& {Swank}}]{bradt93}
{Bradt}, H.~V., {Rothschild}, R.~E., \& {Swank}, J.~H. 1993, A\&AS, 97, 355

\bibitem[{{Butters} {et~al.}(2007){Butters}, {Barlow}, {Norton}, \&
  {Mukai}}]{butters07}
{Butters}, O.~W., {Barlow}, E.~J., {Norton}, A.~J., \& {Mukai}, K. 2007, A\&A,
  475, L29

\bibitem[{{Butters} {et~al.}(2008){Butters}, {Norton}, {Hakala}, {Mukai}, \&
  {Barlow}}]{butters08}
{Butters}, O.~W., {Norton}, A.~J., {Hakala}, P., {Mukai}, K., \& {Barlow},
  E.~J. 2008, A\&A, 487, 271

\bibitem[{{Lehto}(1997)}]{lehto97}
{Lehto}, H.~J. 1997, in Applications of time series analysis in astronomy and
  meteorology, ed. T.~{Subba Rao}, M.~B. {Priestley}, \& O.~{Lessi} (London
  Chapman and Hall)

\bibitem[{{Masetti} {et~al.}(2006){Masetti}, {Morelli}, {Palazzi}, {Galaz},
  {Bassani}, {Bazzano}, {Bird}, {Dean}, {Israel}, {Landi}, {Malizia},
  {Minniti}, {Schiavone}, {Stephen}, {Ubertini}, \& {Walter}}]{masetti06}
{Masetti}, N., {Morelli}, L., {Palazzi}, E., {et~al.} 2006, A\&A, 459, 21

\bibitem[{{Norton} \& {Watson}(1989)}]{norton89}
{Norton}, A.~J. \& {Watson}, M.~G. 1989, MNRAS, 237, 853

\bibitem[{{Parker} {et~al.}(2005){Parker}, {Norton}, \& {Mukai}}]{parker05}
{Parker}, T.~L., {Norton}, A.~J., \& {Mukai}, K. 2005, A\&A, 439, 213

\bibitem[{{Pretorius}(2009)}]{pretorius09}
{Pretorius}, M.~L. 2009, arXiv:0901.2841

\bibitem[{{Warner}(1995)}]{warner95}
{Warner}, B. 1995, {Cataclysmic variable stars} (Cambridge Astrophysics Series,
  Cambridge, New York: Cambridge University Press, |c1995)

\end{thebibliography}

\end{document}